\documentclass[12pt]{article}
\usepackage{amsmath,amssymb,theorem,cite,epsfig,url,psfrag,eepic,mathtools,amsmath,mathtools,graphicx,xcolor,bm,upgreek}
\usepackage[bbgreekl]{mathbbol}
\DeclareSymbolFontAlphabet{\mathbbm}{bbold}
\DeclareSymbolFontAlphabet{\mathbb}{AMSb}
\usepackage{hyperref}
\advance \topmargin by -\headheight
\advance \topmargin by -\headsep     
\evensidemargin \oddsidemargin
\def\Maketitle{{\def\newpage{}\maketitle}}
\makeatletter
\def\Appendix{\appendix
	\def\@seccntformat##1{Appendix~\csname the##1\endcsname.~~}}
\makeatother
\makeatletter
\@addtoreset{equation}{section}

\makeatother
\makeatletter
\newcommand{\zerounderset}[3][\mathord]{%
	#1{\vtop{
			\let\\\cr
			\baselineskip\z@skip\lineskip.25ex
			\ialign{\hidewidth$##$\hidewidth\crcr
				\omit$#3$\cr
				#2\crcr
			}%
	}}%
}
\makeatother
\begin{document}
\rightline{\texttt{\today}}
\title{\textbf{R-matrix formulation of affine Yangian of $\widehat{\mathfrak{gl}}(1|1)$}\vspace*{.3cm}}
\date{}
\author{Dmitry Kolyaskin$^{1,2}$, Alexey Litvinov$^{2,3}$ and Arkady Zhukov$^{1,2}$\\[\medskipamount]
\parbox[t]{0.85\textwidth}{\normalsize\it\centerline{1. Moscow Institute of Physics and Technology, 141700
Dolgoprudny, Russia}}\\
\parbox[t]{0.85\textwidth}{\normalsize\it\centerline{2. Landau Institute for Theoretical Physics, 142432 Chernogolovka, Russia}}\\
\parbox[t]{0.85\textwidth}{\normalsize\it\centerline{3. Center for Advanced Studies, Skolkovo Institute of Science and Technology, 143026 Moscow, Russia}}}
\Maketitle
\begin{abstract}
We study $\mathcal{N}=2$ superconformal field theory and define the R-matrix which acts as an intertwining operator between different realizations of $\mathcal{N}=2$ $W-$algebras of type $A$. Using this R-matrix we define $RLL$ algebra and relate it to current realization of affine Yangian of $\widehat{\mathfrak{gl}}(1|1)$.
\end{abstract}
\section{Introduction}\label{intro}
It has been known for a while that integrability in CFT can be studied using the affine Yangian symmetry. The basic example -- the affine Yangian of $\widehat{\mathfrak{gl}}(1)$, or $Y(\widehat{\mathfrak{gl}}(1))$ for shortness, has been introduced in R-matrix formulation by Maulik and Okounkov \cite{Maulik:2012wi}. The R-matrix of \cite{Maulik:2012wi} is closely related to the Liouville reflection operator \cite{Zamolodchikov:1995aa}, which intertwines two Miura representations for Virasoro algebra. The algebra under the same name has been introduced by Tsymbaliuk in \cite{Tsymbaliuk:2014fvq} by explicit commutation relations (the so called current realization). In \cite{Litvinov:2020zeq} it has been shown that both algebras differ by an infinite-dimensional center.

In \cite{Prochazka:2015deb,Gaberdiel:2017dbk} the relation between $Y(\widehat{\mathfrak{gl}}(1))$ and bosonic $W_{1+\infty}$ algebra has been established. In particular, some of low lying generators of $W_{1+\infty}$ algebra were expressed explicitly in terms of the affine Yangian generators. Later in \cite{Gaberdiel:2017hcn,Gaberdiel:2018nbs}  Gaberdiel et all studied $\mathcal{N}=2$ superconformal version of $W_{1+\infty}$ algebra and defined, what they called "the supersymmetric affine Yangian algebra". The analysis in \cite{Gaberdiel:2017hcn,Gaberdiel:2018nbs} has been using the fact that $\mathcal{N}=2$ $W_{1+\infty}$ algebra contains two commuting bosonic $W_{1+\infty}$ subalgebras. The corresponding Yangian algebra were built  in the so called gluing basis: having two explicitly commuting $Y(\widehat{\mathfrak{gl}}(1))$ algebras and additional fermionic generators, which transform as bi-fundamental representations (see also \cite{Li:2019nna,Li:2019lgd} for later developments).

In \cite{Li:2020rij} (see also \cite{Galakhov:2020vyb,Galakhov:2021xum}) the current realization of $Y(\widehat{\mathfrak{gl}}(1|1))$ has been given\footnote{We note that \cite{Li:2020rij} aimed to construct various Yangian algebras from geometry and contains much more examples. The algebra $Y(\widehat{\mathfrak{gl}}(1|1))$ corresponds to the resolved conifold geometry and the so called pyramid partitions.}. Moreover later in \cite{Galakhov:2021xum} it has been explicitly stated that $Y(\widehat{\mathfrak{gl}}(1|1))$ defined in \cite{Li:2020rij} (together with Serre relations, which appear later in \cite{Galakhov:2021xum}) should correspond to $\mathcal{N}=2$ $W_{1+\infty}$ algebra in the same way as  $Y(\widehat{\mathfrak{gl}}(1))$ corresponds to bosonic  $W_{1+\infty}$ algebra.

In this paper we relate $\mathcal{N}=2$ $W-$algebras and $Y(\widehat{\mathfrak{gl}}(1|1))$. We follow another approach similar to \cite{Litvinov:2020zeq}. Namely, we define the R-matrix, which acts in the tensor product of two typical representations of $\mathfrak{gl}(1|1)$ current algebra, then define the $RLL$ algebra and derive current relations of $Y(\widehat{\mathfrak{gl}}(1|1))$ out of it. This paper is organized as follows. In section \ref{W-algebra} we review $\mathcal{N}=2$ $W_n-$algebras of type $A$ and their free field representations. In section \ref{bcbetagamma} we introduce $\widehat{\mathfrak{gl}}(1|1)$ algebra and its representations through $bc\beta\gamma$-system. In section \ref{R-matrix} we define the $R-$matrix for $\widehat{\mathfrak{gl}}(1|1)$. In sections \ref{RLL} and \ref{current-relations} we relate $RLL$ algebra and current realization of $Y(\widehat{\mathfrak{gl}}(1|1))$. Sections \ref{R-matrix}, \ref{RLL} and \ref{current-relations} constitute main results of our paper. In section \ref{concl} we provide concluding remarks and directions for future studies. 
\section{\texorpdfstring{$\mathcal{N}=2$}{N=2} superconformal \texorpdfstring{$W-$}{W}algebras}\label{W-algebra}
The $\mathcal{N}=2$ $W_{n+1}$-algebras, which play the key role in our study, correspond to $SU(n+1)/U(n)$ Kazama-Suzuki models \cite{Kazama:1988qp}. These algebras contain one $\mathcal{N}=2$ multiplet of holomorphic currents for each integers spin $s=1,\dots,n$. In particular, for $s=1$ one has the currents $(J,G^{+},G^{-},T)$ of spins $(1,\frac{3}{2},\frac{3}{2},2)$, which form  $\mathcal{N}=2$ Virasoro algebra (whose commutation relations we present in the form of OPE's)
\begin{equation}\label{N=2-OPE}
\begin{aligned}
&T(z)T(w)=\frac{c}{2(z-w)^{4}}+\frac{2T(w)}{(z-w)^{2}}+\frac{\partial T(w)}{z-w}+\dots,\\
&T(z)J(w)=\frac{J(w)}{(z-w)^{2}}+\frac{\partial J(w)}{z-w}+\dots,\\
&T(z)G^{\pm}(w)=\frac{3G^{\pm}(w)}{2(z-w)^{2}}+\frac{\partial G^{\pm}(w)}{z-w}+\dots,\\
&J(z)J(w)=\frac{c}{3(z-w)^{2}}+\dots,\\
&J(z)G^{\pm}(w)=\pm\frac{G^{\pm}(w)}{z-w}+\dots,\\
&G^{+}(z)G^{-}(w)=\frac{c}{3(z-w)^{3}}+\frac{J(w)}{(z-w)^{2}}+\frac{T(w)+\frac{1}{2}\partial J(w)}{z-w}+\dots
\end{aligned}
\end{equation}
with other OPE's regular. Commutation relations for higher spin currents become much more involved and are, in principle, not sufficiently illuminating and convenient to handle.

Instead, we will use free-field representation of $\mathcal{N}=2$ $W_{n+1}$-algebras, which is obtained by Hamiltonian reduction of $\widehat{\mathfrak{sl}}(n+1|n)$ \cite{Evans:1990qq,Komata:1990cb,Ito:1991wb}. We work in $\mathcal{N}=1$ formalism and introduce $2n$ real holomorphic superfields $\vec{\boldsymbol{\Phi}}(z,\theta)=(\boldsymbol{\Phi}_1(z,\theta),\dots,\boldsymbol{\Phi}_{2n}(z,\theta))$
\begin{equation}
\begin{gathered}
    \boldsymbol{\Phi}_k(z,\theta)\overset{\text{def}}{=}\Phi_k(z)+\theta\chi_k(z),\\
    \Phi_i(z)\Phi_j(w)=-\frac{\delta_{ij}}{2}\log(z-w)+\dots,\quad
    \chi_i(z)\chi_j(w)=\frac{\delta_{ij}}{2}\frac{1}{z-w}+\dots
\end{gathered}
\end{equation}
Let  $\vec{\alpha}_r=(\alpha_{r,1},\dots,\alpha_{r,2n})$ with $r=1,\dots,2n$ be the simple fermionic roots of $\mathfrak{sl}(n+1|n)$ with the Gram matrix
\begin{equation}
    \big(\vec{\alpha}_r\cdot\vec{\alpha}_s\big)=
    \begin{pmatrix}
        0&-1&0&\cdots&\cdots\\
        -1&0&1&\cdots&\cdots\\
        0&1&0&\cdots&\cdots\\
        \cdots&\cdots&\cdots&\cdots&\cdots\\
        \cdots&\cdots&\cdots&\cdots&\cdots
    \end{pmatrix},
\end{equation}
$\vec{\omega}_r$ being the dual basis $(\vec{\omega}_r\cdot\vec{\alpha}_s)=\delta_{rs}$ and 
\begin{equation}
      \vec{h}_r\overset{\text{def}}{=}(-1)^r\left(\vec{\omega}_r-\vec{\omega}_{r-1}\right)
      \quad\text{for}\quad r=1,\dots,2n+1,
\end{equation}
where $\vec{\omega}_0=\vec{\omega}_{2n+1}=0$. Then  $\mathcal{N}=2$ $W_{n+1}$-algebra is defined by \cite{Evans:1990qq,Komata:1990cb,Ito:1991wb}
\begin{equation}\label{Miura-map-1}
\begin{gathered}
    :\left(b^{-1}D-D\big(\vec{h}_1\cdot\vec{\boldsymbol{\Phi}}(z,\theta)\big)\right)
    \left(b^{-1}D-D\big(\vec{h}_2\cdot\vec{\boldsymbol{\Phi}}(z,\theta)\big)\right)\dots
    \left(b^{-1}D-D\big(\vec{h}_{2n+1}\cdot\vec{\boldsymbol{\Phi}}(z,\theta)\big)\right):=\\
    =
    \left(b^{-1}D\right)^{2n+1}+\boldsymbol{W}_{\frac{1}{2}}(z,\theta)\left(b^{-1}D\right)^{2n}+\boldsymbol{W}_1(z,\theta)\left(b^{-1}D\right)^{2n-1}+
    \boldsymbol{W}_\frac{3}{2}(z,\theta)\left(b^{-1}D\right)^{2n-2}+\dots
\end{gathered}
\end{equation}
where
\begin{equation}
    D=\frac{\partial}{\partial\theta}-\theta\partial.
\end{equation}
We have $\boldsymbol{W}_{\frac{1}{2}}(z,\theta)=0$,
\begin{equation}\label{W-W-modes}
    \boldsymbol{W}_{1}(z,\theta)=J(z)+i\theta(G^{+}(z)-G^{-}(z)),\quad
    b\boldsymbol{W}_{\frac{3}{2}}(z,\theta)=iG^{-}(z)+\theta\left(T(z)+\frac{1}{2}\partial J(z)\right)
\end{equation}
etc. The currents \eqref{W-W-modes} satisfy commutation relations of $\mathcal{N}=2$ Virasoro algebra \eqref{N=2-OPE} with the central charge
\begin{equation}
      c=3n\left(1+\frac{n+1}{b^2}\right).
\end{equation}
Higher currents in \eqref{Miura-map-1} provide its extension.

In fact, it is more convenient to work in $\mathfrak{gl}(n+1|n)$ rather than in $\mathfrak{sl}(n+1|n)$ terms. In order to do it, we extend our field as $\vec{\boldsymbol{\Phi}}(z,\theta)=(\boldsymbol{\Phi}_0(z,\theta),\boldsymbol{\Phi}_1(z,\theta),\dots,\boldsymbol{\Phi}_{2n}(z,\theta))$. In this case we can choose simple roots as
\begin{equation}
    \left(\vec{\alpha}_{2r+1}\cdot\vec{\boldsymbol{\Phi}}(z,\theta)\right)=
    i\boldsymbol{\Phi}_{2r}(z,\theta)-\boldsymbol{\Phi}_{2r+1}(z,\theta),\quad
    \left(\vec{\alpha}_{2r}\cdot\vec{\boldsymbol{\Phi}}(z,\theta)\right)=
    \boldsymbol{\Phi}_{2r-1}(z,\theta)-i\boldsymbol{\Phi}_{2r}(z,\theta).
\end{equation}
In this case the map \eqref{Miura-map-1} takes very simple form
\begin{equation}\label{Miura-map-2}
    \left(b^{-1}D-iD\boldsymbol{\Phi}_0(z,\theta)\right)
    \left(b^{-1}D-D\boldsymbol{\Phi}_1(z,\theta)\right)\dots
    \left(b^{-1}D-iD\boldsymbol{\Phi}_{2n}(z,\theta)\right)
\end{equation}
We denote the corresponding algebra as $W_{\mathfrak{gl}(n+1|n)}$. Similarly $W_{\mathfrak{gl}(n|n)}$ corresponds to \eqref{Miura-map-2} with the factor $\left(b^{-1}D-iD\boldsymbol{\Phi}_0(z,\theta)\right)$ dropped.

It is important to note that both algebras $W_{\mathfrak{gl}(n+1|n)}$ and $W_{\mathfrak{gl}(n|n)}$ can be defined as commutants of certain set of  screening operators \cite{Ito:1991wb}. For $W_{\mathfrak{gl}(n|n)}$ it will be convenient to use complex fields ($k=1,\dots,n$)
\begin{equation}
\begin{aligned}
   & \boldsymbol{X}_k(z,\theta)\overset{\text{def}}{=}\boldsymbol{\Phi}_{2k-1}(z,\theta)+i\boldsymbol{\Phi}_{2k}(z,\theta)=X_k(z)+\theta\psi_k(z),\\
   & \boldsymbol{X}^*_k(z,\theta)\overset{\text{def}}{=}\boldsymbol{\Phi}_{2k-1}(z,\theta)-i\boldsymbol{\Phi}_{2k}(z,\theta)=X_k^*(z)+\theta\psi_k^*(z),
\end{aligned}
\end{equation}
where $(X_k(z),X_k^*(z),\psi(z),\psi^*(z))$ have the following OPE's
\begin{equation}
    X_i(z)X_j^*(w)=-\delta_{ij}\log(z+w)+\dots,\quad
    \psi_i(z)\psi_j^*(w)=\frac{\delta_{ij}}{z-w}+\dots
\end{equation}
In these terms the screening operators for $W_{\mathfrak{gl}(n|n)}$ algebra $(S_1,S_{1,2},S_2,S_{2,3},\dots,S_{n-1,n},S_n)$ have the form \cite{Ito:1991wb}
\begin{equation}\label{Screenings}
\begin{aligned}
    &S_{k}=b^{-1}\oint e^{b\big(\vec{\alpha}_{2k-1}\cdot\vec{\boldsymbol{\Phi}}(z,\theta)\big)}dzd\theta=b^{-1}\oint e^{b\boldsymbol{X}_k^*(z,\theta)}dzd\theta=\oint\psi^*_k(z)e^{bX^*_k(z)}dz,
    \\
    &S_{k,k+1}=2b^{-1}\oint e^{b\big(\vec{\alpha}_{2k}\cdot\vec{\boldsymbol{\Phi}}(z,\theta)\big)}dzd\theta=2b^{-1}
    \oint e^{\frac{b}{2}\left(\boldsymbol{X}_k(z,\theta)-\boldsymbol{X}_{k+1}(z,\theta)
    -\boldsymbol{X}^*_k(z,\theta)-\boldsymbol{X}^*_{k+1}(z,\theta)\right)}dzd\theta=\\
    &=\oint e^{\frac{b}{2}\left(X_k(z)-X_{k+1}(z)
    -X^*_k(z)-X^*_{k+1}(z)\right)}\left(\psi_k(z)-\psi_{k+1}(z)
    -\psi^*_k(z)-\psi^*_{k+1}(z)\right)dz.
\end{aligned}
\end{equation}
We note that all screening fields \eqref{Screenings} are fermionic ones. There are also bosonic screening fields \cite{Ito:1991wb}. 
\section{\texorpdfstring{$\mathfrak{gl}(1|1)$}{gl(1|1)} current algebra and \texorpdfstring{$bc\beta\gamma$}{bcbetagamma}-system}\label{bcbetagamma}
The commutation relations of $\widehat{\mathfrak{gl}}(1|1)$ algebra \cite{Rozansky:1992rx}  can be written in  terms of OPE of two bosonic currents $J^{E}(z)$, $J^{N}(z)$ and two fermionic currents $J^{\pm}(z)$:
\begin{equation}
    \begin{aligned}
       &J^{N}(z)J^{E}(w)=\frac{\kappa}{(z-w)^2}+\dots\\
       &J^{+}(z)J^{-}(w)=\frac{\kappa}{(z-w)^2}+\frac{J^E(w)}{(z-w)^2}+\dots\\
       &J^{\pm}(z)J^{N}(w)=\mp\frac{J^{\pm}(w)}{z-w}+\dots
    \end{aligned}
\end{equation}
The parameter $\kappa$ called the level can be set $\kappa=1$ by the simple rescaling $J^E(z)\rightarrow\kappa J^E(z)$ and $J^+(z)\rightarrow\kappa J^+(z)$. It implies that $\widehat{\mathfrak{gl}}(1|1)$ algebra admits free field representation \cite{Witten:1983ar} via the $bc\beta\gamma$-system. Namely, $bc\beta\gamma$-system is the system of complex fermions $\alpha(z)$, $\alpha^*(z)$ and complex bosons $\mathrm{a}(z)$, $\mathrm{a}^*(z)$ with OPE's (other OPE's are trivial)
\begin{equation}
    \alpha(z)\alpha^*(w)=\frac{1}{z-w}+\dots,\qquad  \mathrm{a}(z)\mathrm{a}^*(w)=\frac{1}{z-w}+\dots
\end{equation}
Then one can realize the currents $J^{E}(z)$, $J^{N}(z)$ and $J^{\pm}(z)$ as bilinears (Wick normal ordering is assumed)
\begin{equation}
    \boldsymbol{E}\overset{\text{def}}{=}
    \begin{pmatrix}
       \frac{J^E}{2}+J^N&J^+\\
       J^-&\frac{J^E}{2}-J^N
    \end{pmatrix}=
    \begin{pmatrix}
       \alpha^*\alpha&\alpha^*\mathrm{a}\\
       \mathrm{a}^*\alpha&\mathrm{a}^*\mathrm{a}
    \end{pmatrix}
\end{equation}

For our purposes it will be convenient to further realize $(\alpha(z),\alpha^*(z),\mathrm{a}(z),\mathrm{a}^*(z))$ by the complex superfield $\boldsymbol{X}(z,\theta)=X(z)+\theta\psi(z)$, $\boldsymbol{X}^*(z,\theta)=X^*(z)+\theta\psi^*(z),$
as the Wick ordered fields
\begin{equation}\label{bcbetagamma-bosonization}
    \begin{aligned}
       & \alpha=\psi e^{-\frac{X}{b}},\quad&& \mathrm{a}=e^{-\frac{X}{b}},\\
       & \alpha^*=\psi^* e^{\frac{X}{b}},\quad&& \mathrm{a}^*=\left(\psi\psi^*-b\partial X^*\right)e^{\frac{X}{b}}.
    \end{aligned}
\end{equation}
Here $b$ is an arbitrary parameter, which can be re-scaled out by the transformation $X\rightarrow bX$, $X^*\rightarrow b^{-1}X^*$, ho we will keep it as it is. It is important to note that the currents \eqref{bcbetagamma-bosonization} can be equivalently defined by the following Wakimoto type screening operator\footnote{We remind that $\widehat{\mathfrak{gl}}(2)_{\kappa}$ algebra admits free field representation in terms of two bosonic fields $\varphi_1(z)$, $\varphi_2(z)$ and $\beta\gamma$-system realized by Wakimoto screening operator \cite{Wakimoto:1986gf}
\begin{equation}
   \mathcal{S}=\oint e^{b(\varphi_1(z)-\varphi_2(z))}\beta(z)dz,   
\end{equation}
with $\kappa=-2-b^{-2}$. In $\widehat{\mathfrak{gl}}(1|1)$ case $(\varphi_1(z)-\varphi_2(z))$ is replaced by $X^*(z)$ and $\beta(z)$ by $\psi^*(z)$.
}
\begin{equation}\label{bcbetagamma-screening}
    \mathcal{S}=b^{-1}\oint e^{b\boldsymbol{X}^*(z,\theta)}dzd\theta=\oint\psi^*(z)e^{bX^*(z)}dz.
\end{equation}
We note also that \eqref{bcbetagamma-bosonization} admit the following simple superfield expression
\begin{equation}
    \mathrm{a}(z)-\frac{\theta}{b}\alpha(z)=e^{-\frac{\boldsymbol{X}(z,\theta)}{b}},\quad
    \alpha^*(z)+\frac{\theta}{b}\mathrm{a}^*(z)=D\boldsymbol{X}^*(z,\theta)e^{\frac{\boldsymbol{X}(z,\theta)}{b}}.
\end{equation}

In terms of the fields $(X,X^*,\psi,\psi^*)$  the representation of $\widehat{\mathfrak{gl}}(1|1)$ takes the form
\begin{equation}
    \boldsymbol{E}=
    \begin{pmatrix}
      b^{-1}\partial X-\psi\psi^*&\psi^*\\
      \partial\psi-b\psi\partial X^*&\psi\psi^*-b^{-1}\partial X-b\partial X^*
    \end{pmatrix}
\end{equation}
In these notes we consider the so called typical (or generic) representations of $\widehat{\mathfrak{gl}}(1|1)$. One distinguishes between NS and R modules. We consider NS module for simplicity. We define the mode expansion of the currents on the cylinder $z=e^{-ix}$: $x\sim x+2\pi$
\begin{equation}
\begin{gathered}
    \partial X(x)=\sum_{k\in\mathbb{Z}}X_ke^{-ikx},\quad
    \partial X^*(x)=\sum_{k\in\mathbb{Z}}X^*_ke^{-ikx},\\
    \psi(x)=i^{\frac{1}{2}}\sum_{r\in\mathbb{Z}+\frac{1}{2}}\psi_re^{-irx},\quad
    \psi^*(x)=i^{\frac{1}{2}}\sum_{r\in\mathbb{Z}+\frac{1}{2}}\psi^*_re^{-irx}.
\end{gathered}
\end{equation}
with commutation relations (other commutation relations are trivial)
\begin{equation}
    [X_m,X_n^*]=m\delta_{m,-n},\quad\{\psi_{r},\psi_{s}^*\}=\delta_{r,-s}.
\end{equation}
Then the Fock module $\mathcal{F}_{\boldsymbol{u}}$, where $\boldsymbol{u}=(u,u^*)$, is generated from  the vacuum state $|\boldsymbol{u}\rangle$
\begin{equation}
\begin{gathered}
    X_{n}|\boldsymbol{u}\rangle=X^*_{n}|\boldsymbol{u}\rangle=\psi_{r}|\boldsymbol{u}\rangle=
    \psi^*_{r}|\boldsymbol{u}\rangle=0\quad\text{for}\quad n>0,\,r>0,\\
    X_{0}|\boldsymbol{u}\rangle=-iu|\boldsymbol{u}\rangle,\quad
    X_{0}^*|\boldsymbol{u}\rangle=-iu^*|\boldsymbol{u}\rangle,
\end{gathered}
\end{equation}
by the action of the creation operators
\begin{equation}\label{NS-module}
    \mathcal{F}_{\boldsymbol{u}}=\textrm{Span}\left(\psi_{-\boldsymbol{s}^*}^*\psi_{-\boldsymbol{s}}X^*_{-\boldsymbol{\lambda}^*}X_{-\boldsymbol{\lambda}}|\boldsymbol{u}\rangle\right),\quad\text{where}\quad
    X_{-\boldsymbol{\lambda}}=X_{-\lambda_1}X_{-\lambda_2}\dots\,\text{etc}.,
\end{equation}
where $\boldsymbol{\lambda}$ and $\boldsymbol{\lambda}^*$ are partitions
\begin{equation}
    \boldsymbol{\lambda}=\{\lambda_1\geq\lambda_2\geq\dots\},\qquad
    \boldsymbol{\lambda}^*=\{\lambda_1^*\geq\lambda_2^*\geq\dots\},
\end{equation}
and $\boldsymbol{s}$ and $\boldsymbol{s}^*$ are strict partitions (fermionic partitions)
\begin{equation}
    \boldsymbol{s}=\{s_1>s_2>\dots\},\qquad
    \boldsymbol{s}^*=\{s_1^*>s_2^*>\dots\},
\end{equation}

The character of $\mathcal{F}_{\boldsymbol{u}}$ has the form
\begin{equation}\label{character}
    \chi(q,t)\overset{\text{def}}{=}\textrm{Tr}(q^{L_0}t^{c_0})\Bigl|_{\mathcal{F}_{\boldsymbol{u}}}=\prod_{k=1}^{\infty}\frac{\left(1+tq^{k-\frac{1}{2}}\right)\left(1+t^{-1}q^{k-\frac{1}{2}}\right)}{(1-q^k)^2},
\end{equation}
where
\begin{equation}
    L_0=\sum_{k>0}\left(X_{-k}X_{k}^*+X_{-k}^*X_{k}\right)+
    \sum_{s>0}s\left(\psi_{-s}\psi^*_s+\psi_{-s}^*\psi_s\right),\quad
    c_0=\sum_{s>0}\left(\psi_{-s}\psi^*_s-\psi_{-s}^*\psi_s\right).
\end{equation}
\section{R-matrix}\label{R-matrix}
In order to define the R-matrix for $\widehat{\mathfrak{gl}}(1|1)$, we take two copies of $\widehat{\mathfrak{gl}}(1|1)$ algebra defined by two screening fields
\begin{equation}
    \mathcal{S}_1=\int\psi_1^*(z)e^{bX_1^*(z)}dz\quad\text{and}\quad
    \mathcal{S}_2=\int\psi_2^*(z)e^{bX_2^*(z)}dz
\end{equation}
and define the additional screening field
\begin{equation}\label{S12-tilde}
    \tilde{\mathcal{S}}_{1,2}\overset{\text{def}}{=}\int (\psi_1(z)-\psi_2(z))e^{\frac{b}{2}\left(X_1(z)-X_2(z)-X_1^*(z)-X_2^*(z)\right)}dz,
\end{equation}
which commutes with the diagonal $\widehat{\mathfrak{gl}}(1|1)$ current $\boldsymbol{E}_1+\boldsymbol{E}_2$.

The commutant of this additional screening is actually larger and can be associated with the $W(\mathfrak{gl}(2|2))$ algebra introduced in section \ref{W-algebra}. We note that the screening operator $\tilde{\mathcal{S}}_{1,2}$ given by \eqref{S12-tilde} differs from $S_{1,2}$ given by \eqref{Screenings} by the replacement in the prefactor
\begin{equation}
    \psi_1(z)-\psi_2(z)\rightarrow\psi_1(z)-\psi_2(z)-\psi_1^*(z)-\psi_2^*(z).
\end{equation}
However two algebras defined by $(\mathcal{S}_1,\mathcal{S}_{1,2},\mathcal{S}_2)$ and by $(\mathcal{S}_1,\tilde{\mathcal{S}}_{1,2},\mathcal{S}_2)$ are actually isomorphic and are related by marginal deformations. It can be seen as follows. Consider the Lax operator
\begin{equation}
    \mathcal{L}=\left(b^{-1}D-D\boldsymbol{\Phi}_1\right)\left(b^{-1}D-iD\boldsymbol{\Phi}_2\right)=\left(b^{-1}D\right)^2-D\boldsymbol{X}^*b^{-1}D+\frac{1}{2}D\boldsymbol{X}^*D\boldsymbol{X}+\frac{1}{2b}D^2\left(\boldsymbol{X}^*-\boldsymbol{X}\right),
\end{equation}
where $\boldsymbol{X}=\boldsymbol{\Phi}_1+i\boldsymbol{\Phi}_2$ and perform the formal scale transformation
\begin{equation}
    \psi\rightarrow \lambda^{-\frac{1}{2}}\psi,\quad
    \psi^*\rightarrow \lambda^{\frac{1}{2}}\psi^*,\quad
    \theta\rightarrow \lambda^{\frac{1}{2}}\theta,\quad
    \frac{\partial}{\partial\theta}\rightarrow \lambda^{-\frac{1}{2}}\frac{\partial}{\partial\theta},
\end{equation}
which does not change commutation relations. Then one has
\begin{equation}
  \mathcal{L}=\mathcal{L}^{(0)}+\lambda\mathcal{L}^{(1)},
\end{equation}
where (we use $D^2=-\partial$)
\begin{equation}
    \mathcal{L}^{(0)}=-b^{-2}\partial+\frac{1}{2}\left(b^{-1}\partial X-\psi\psi^*-b^{-1}\partial X^*\right)+
    \frac{1}{2b}\left(\partial\psi-b\psi\partial X^*\right)\theta
    -b^{-1}\psi^*\frac{\partial}{\partial\theta}+b^{-1}\partial X^*\theta\frac{\partial}{\partial\theta},
\end{equation}
and
\begin{equation}
    \mathcal{L}^{(1)}=\left(b^{-1}\psi^*\partial+\frac{1}{2}\psi^*\partial X-\frac{1}{2b}\partial\psi^*\right)\theta.
\end{equation}
It is clear that if $\mathcal{L}_1\cdot\mathcal{L}_2$ "commutes" with the screening charges $(\mathcal{S}_1,\mathcal{S}_{1,2},\mathcal{S}_{2})$ then $\mathcal{L}_1^{(0)}\cdot\mathcal{L}_2^{(0)}$ "commutes" with $(\mathcal{S}_1,\tilde{\mathcal{S}}_{1,2},\mathcal{S}_{2})$\footnote{The current $A(z)$ is said to "commute" with the screening field $\mathcal{S}=\int\mathcal{V}(\xi)d\xi$ if
\begin{equation}
  \oint_{\mathcal{C}_z}\mathcal{V}(\xi)A(z)d\xi=0.
\end{equation}.}.

Having defined the Lax operator $\mathcal{L}^{(0)}$,  we define the $R-$matrix as an intertwining operator
\begin{equation}\label{RLL-Miura}
    \mathcal{R}_{ij}\mathcal{L}^{(0)}_i\mathcal{L}^{(0)}_j=
    \mathcal{L}^{(0)}_j\mathcal{L}^{(0)}_i\mathcal{R}_{ij},
\end{equation}
where $\mathcal{L}^{(0)}_j\overset{\text{def}}{=}\mathcal{L}^{(0)}(X_i,X_i^*,\psi_i,\psi_i^*)$. We note that \eqref{RLL-Miura} automatically implies the Yang-Baxter equation
\begin{equation}\label{YB-equation}
    \mathcal{R}_{12}\mathcal{R}_{13}\mathcal{R}_{23}=\mathcal{R}_{23}\mathcal{R}_{13}\mathcal{R}_{12}.
\end{equation}

The $W_{\mathfrak{gl}(2|2)}$ algebra defined by $\mathcal{L}_1^{(0)}\mathcal{L}_2^{(0)}$ is obtained from the one defined by $\mathcal{L}_1\mathcal{L}_2$ in the limit $\lambda\rightarrow0$. It consists of eight currents
\begin{equation}
  \begin{gathered}    
    \Psi_{12}^*\overset{\text{def}}{=}\psi_1^*+\psi_2^*,
    \quad U_{12}\overset{\text{def}}{=}\partial X_1^*+\partial X_2^*,
    \quad
    J_{12}\overset{\text{def}}{=}\psi_1\psi_1^*+\psi_2\psi_2^*-
    \frac{1}{b}\left(\partial X_1+\partial X_2\right),\\
    G_{12}\overset{\text{def}}{=}i\left(\psi_1\partial X_1^*+\psi_2\partial X_2^*-\frac{1}{b}\left(\partial\psi_1+\partial\psi_2\right)\right),\\
   G_{12}^*\overset{\text{def}}{=}i\left(\psi_{1}^*(\partial X_{1}-\partial X_{2}^*)+\psi_{2}^*(\partial X_{2}+\partial X_{1}^*)-\frac{1}{b}\left(\partial\psi_{1}^*-\partial\psi_2^*\right)\right),\\
    T_{12}\overset{\text{def}}{=}-\partial X_1 \partial X_{1}^*-\partial X_2 \partial X_{2}^*+\dfrac{1}{b}(\partial^2X_1^*-\partial^2X_2^*)+\partial\psi_1\psi_1^*+\partial\psi_2\psi_2^*-\dfrac{1}{2}\partial J_{12},
  \end{gathered}
\end{equation}
\begin{multline}
    \Theta_{12}\overset{\text{def}}{=}-\dfrac{1}{8b^2}((\partial X_1-\partial X^*_1)^2+(\partial X_2-\partial X^*_2)^2)-\dfrac{1}{2b^3}(\partial^2 X_2-\partial^2 X_2^*)+\dfrac{1}{4b}(\partial X_1-\partial X_1^*)\psi_1\psi_1^*+\\+\dfrac{1}{4b}(\partial X_2-\partial X_2^*)\psi_2\psi_2^*+\dfrac{1}{2b}\partial X_2^*\psi_1^*\psi_2+\dfrac{1}{8}(\psi_1^*\partial\psi_1+\psi_1\partial\psi_1^*+\psi_2^*\partial\psi_2+\psi_2\partial\psi_2^*)+\\+\dfrac{1}{2b^2}(\psi_2\partial\psi_2^*++\partial\psi_2\psi_2^*-\psi_1^*\partial\psi_2),
\end{multline}
and
\begin{multline}
    H_{12}\overset{\text{def}}{=}\dfrac{1}{4}\psi_1\psi_1^*\psi_2\partial X_2^*+\dfrac{1}{4}\psi_2\psi_2^*\psi_1\partial X_1^*-\\-
    \frac{1}{4b}\left(\big(\partial X_1+\partial X_1^*\big)\partial X_2^*\psi_2+\big(\partial X_2-\partial X_2^*\big)\partial X_1^*\psi_1+\psi_1\psi_1^*\partial\psi_2+\psi_2\psi_2^*\partial\psi_1\right)+\\+
    \frac{1}{4b^2}\left(\big(\partial X_1+\partial X_1^*+2\partial X_2^*\big)\partial\psi_2+\big(\partial X_2-\partial X_2^*\big)\partial\psi_1+2\psi_2\partial^2X_2^*\right)-\frac{1}{2b^3}\partial^2\psi_2,
\end{multline}
where all the densities are \emph{Wick ordered}. We note that $\left(J_{12},G_{12},G_{12}^*,T_{12}\right)$ form $\mathcal{N}=2$ Virasoro algebra with the central charge
\begin{equation}
    c=6.
\end{equation}
At the same time the currents $(\Psi^*_{12},U_{12},J_{12},G_{12})$ correspond to the diagonal current $\boldsymbol{E}_1+\boldsymbol{E}_2$.

Working in a particular representation, the relations \eqref{RLL-Miura} can be used to define the matrix of $R_{12}$. We consider the tensor product of two NS representations \eqref{NS-module}: $\mathcal{F}_{\boldsymbol{u}}\otimes \mathcal{F}_{\boldsymbol{v}}$ with the vacuum state
\begin{equation}
    |\varnothing\rangle\overset{\text{def}}{=}|\boldsymbol{u}\rangle\otimes|\boldsymbol{v}\rangle.
\end{equation}
Since the normalization of $R_{12}$ is not fixed from \eqref{RLL-Miura}, it can be chosen at will. We fix it by $\mathcal{R}_{12}|\varnothing\rangle=|\varnothing\rangle$. Then using  \eqref{RLL-Miura} one can find the action of $\mathcal{R}_{12}$ on any state in $\mathcal{F}_{\boldsymbol{u}}\otimes \mathcal{F}_{\boldsymbol{v}}$.
In particular, using two equations
\begin{equation}
    \mathcal{R}_{12}\left(\Psi_{12}^*\right)_{-\frac{1}{2}}|\varnothing\rangle=\left(\Psi_{21}^*\right)_{-\frac{1}{2}}|\varnothing\rangle,\qquad
    \mathcal{R}_{12}\left(G_{12}^*\right)_{-\frac{1}{2}}|\varnothing\rangle=\left(G_{21}^*\right)_{-\frac{1}{2}}|\varnothing\rangle,
\end{equation}
we find that
\begin{equation}\label{RLL-1}
  \mathcal{R}_{12}\big(\psi_1^*\big)_{-\frac{1}{2}}|\varnothing\rangle=
  \frac{(u-v)-(u^*-v^*)}{(u-v)-(u^*+v^*)+b^{-1}}\big(\psi_1^*\big)_{-\frac{1}{2}}|\varnothing\rangle-
  \frac{2u^*-b^{-1}}{(u-v)-(u^*+v^*)+b^{-1}}\big(\psi_2^*\big)_{-\frac{1}{2}}|\varnothing\rangle,
\end{equation}
and 
\begin{equation}\label{RLL-2}
  \mathcal{R}_{12}\big(\psi_2^*\big)_{-\frac{1}{2}}|\varnothing\rangle=
  -\frac{2v^*-b^{-1}}{(u-v)-(u^*+v^*)+b^{-1}}\big(\psi_1^*\big)_{-\frac{1}{2}}|\varnothing\rangle+
  \frac{(u-v)+(u^*-v^*)}{(u-v)-(u^*+v^*)+b^{-1}}\big(\psi_2^*\big)_{-\frac{1}{2}}|\varnothing\rangle.
\end{equation}
Similarly, using
\begin{equation}
    \mathcal{R}_{12}\left(G_{12}\right)_{-\frac{1}{2}}|\varnothing\rangle=\left(G_{21}\right)_{-\frac{1}{2}}|\varnothing\rangle,\qquad
    \mathcal{R}_{12}\left(H_{12}\right)_{-\frac{1}{2}}|\varnothing\rangle=\left(H_{21}\right)_{-\frac{1}{2}}|\varnothing\rangle,
\end{equation}
we find that
\begin{equation}\label{RLL-3}
  \mathcal{R}_{12}\big(\psi_1\big)_{-\frac{1}{2}}|\varnothing\rangle=
  \frac{(u-v)+(u^*-v^*)}{(u-v)+(u^*+v^*)+b^{-1}}\big(\psi_1\big)_{-\frac{1}{2}}|\varnothing\rangle+
  \frac{2v^*+b^{-1}}{(u-v)+(u^*+v^*)+b^{-1}}\big(\psi_2\big)_{-\frac{1}{2}}|\varnothing\rangle,
\end{equation}
and
\begin{equation}\label{RLL-4}
  \mathcal{R}_{12}\big(\psi_2\big)_{-\frac{1}{2}}|\varnothing\rangle=\frac{2u^*+b^{-1}}{(u-v)+(u^*+v^*)+b^{-1}}\big(\psi_1\big)_{-\frac{1}{2}}|\varnothing\rangle+
  \frac{(u-v)-(u^*-v^*)}{(u-v)+(u^*+v^*)+b^{-1}}\big(\psi_2\big)_{-\frac{1}{2}}|\varnothing\rangle.
\end{equation}
Proceeding in this way, one can compute the matrix of $\mathcal{R}_{12}$ at any given order.

In general, it is clear that the matrix elements of $\mathcal{R}_{12}$ are some rational functions of three parameters
\begin{equation}
    u-v,\quad u^*\quad\text{and}\quad v^*.
\end{equation}
Indeed, from the relation $[\mathcal{R}_{12},U_{12}]=0$ it follows that $\mathcal{R}_{12}$ does not depend on $u+v$. Generic expression for $\mathcal{R}_{12}$ is unlikely can be obtained in a closed form. However, some particular cases and expansions might be useful. In particular, following \cite{Litvinov:2020zeq,Chistyakova:2021yyd}, one can develop the expansion of $\mathcal{R}_{12}(\boldsymbol{u},\boldsymbol{v})$ at large $u-v$.
\section{\texorpdfstring{$RLL$}{RLL} algebra}\label{RLL}
In this section we study $RLL$ algebra provided by the $R-$matrix defined in section \ref{R-matrix}
\begin{equation}\label{RLL-algebra}
    \mathcal{R}_{12}(\boldsymbol{u},\boldsymbol{v})\mathcal{L}_1(\boldsymbol{u})\mathcal{L}_2(\boldsymbol{v})=\mathcal{L}_2(\boldsymbol{v})\mathcal{L}_1(\boldsymbol{u})\mathcal{R}_{12}(\boldsymbol{u},\boldsymbol{v}),
\end{equation}
where as usual $\mathcal{L}_1(\boldsymbol{u})$ and $\mathcal{L}_2(\boldsymbol{v})$ are treated as  matrices acting in  $\widehat{gl}(1|1)$ modules $\mathcal{F}_{\boldsymbol{u}}$ and $\mathcal{F}_{\boldsymbol{v}}$ correspondingly (see \eqref{NS-module}) those entries are operators acting in some unspecified space (called quantum space). Then \eqref{RLL-algebra} is equivalent to the infinite set of quadratic relations between the matrix elements
\begin{equation}
    \langle l|\mathcal{L}(u)|k\rangle,
\end{equation}
where $|k\rangle$ and $|l\rangle$ are arbitrary states from $\mathcal{F}_{\boldsymbol{u}}$. 

However, for various reasons it is more convenient to use the so called current realization of the same algebra. The isomorphism between the $R-$matrix and current realizations is well understood in the case of classical algebras (see e.g. \cite{Jing:2019kws}). For affine algebras it has been studied in \cite{Litvinov:2020zeq} for $\widehat{\mathfrak{gl}}(1)$ and in \cite{Chistyakova:2021yyd} for $\widehat{\mathfrak{gl}}(2)$ cases. 

We will proceed exactly as in  \cite{Litvinov:2020zeq,Chistyakova:2021yyd}. We note however, that there is a conceptual difference compared to $\widehat{\mathfrak{gl}}(1)$ and $\widehat{\mathfrak{gl}}(2)$ cases.  Namely, the R-matrix constructed in section \ref{R-matrix} is rather unusual since unlike most other known R-matrices it cannot be written as a function of the difference of two spectral parameters. The well known example of non-difference R-matrix has been found by Shastry \cite{Shastry:1986zz} and depends on two spectral parameters individually rather than on their difference. Our R-matrix depends on three combinations of four spectral parameters. Nevertheless,  we will show that the relations \eqref{RLL-algebra} imply that some of the currents  are functions of only one spectral parameter, while the others are expressed in terms of them.

This phenomenon can be demonstrated at lower levels.  Let us define the following currents
\begin{equation}\label{ee-ff-currents}
\begin{gathered}
    h(\boldsymbol{u})\overset{\text{def}}{=}\langle\varnothing|\mathcal{L}(\boldsymbol{u})|\varnothing\rangle,\quad
    e(\boldsymbol{u})\overset{\text{def}}{=}h^{-1}(\boldsymbol{u})\langle\varnothing|\mathcal{L}(\boldsymbol{u})\psi_{-\frac{1}{2}}|\varnothing\rangle,\quad
    f(\boldsymbol{u})\overset{\text{def}}{=}\langle\varnothing|\psi_{\frac{1}{2}}\mathcal{L}(\boldsymbol{u})|\varnothing\rangle h^{-1}(\boldsymbol{u}),\\
    e^*(\boldsymbol{u})\overset{\text{def}}{=}\frac{1}{2u^*-b^{-1}}
    h^{-1}(\boldsymbol{u})\langle\varnothing|\mathcal{L}(\boldsymbol{u})\psi^*_{-\frac{1}{2}}|\varnothing\rangle,\quad
    f^*(\boldsymbol{u})\overset{\text{def}}{=}\frac{1}{2u^*+b^{-1}}
    \langle\varnothing|\psi_{\frac{1}{2}}^*\mathcal{L}(\boldsymbol{u})|\varnothing\rangle h^{-1}(\boldsymbol{u}).
\end{gathered}    
\end{equation}
Taking vacuum average of $RLL=LLR$ relation, one finds
\begin{equation}
    [h(\boldsymbol{u}),h(\boldsymbol{v})]=0.
\end{equation}
Similarly, using \eqref{RLL-1}, \eqref{RLL-2}, \eqref{RLL-3} and \eqref{RLL-4}, we find
\begin{equation}\label{he-relation}
    \big((u+u^*)-(v-v^*)+b^{-1}\big)h(\boldsymbol{u})e(\boldsymbol{v})=\big((u-u^*)-(v-v^*)\big)\,e(\boldsymbol{v})h(\boldsymbol{u})\textcolor{blue}{+\big(2u^*+b^{-1}\big)h(\boldsymbol{u})e(\boldsymbol{u})},
\end{equation}
\begin{equation}\label{he*-relation}
    \big((u-u^*)-(v+v^*)+b^{-1}\big)h(\boldsymbol{u})e^*(\boldsymbol{v})=\big((u+u^*)-(v+v^*)\big)\,e^*(\boldsymbol{v})h(\boldsymbol{u})
    \textcolor{blue}{-\big(2u^*-b^{-1}\big)h(\boldsymbol{u})e^*(\boldsymbol{u})},
\end{equation}
\begin{equation}\label{hf-relation}
    \big((u-u^*)-(v+v^*)+b^{-1}\big)f(\boldsymbol{v})h(\boldsymbol{u})=
    \big((u+u^*)-(v+v^*)\big)\,h(\boldsymbol{u})f(\boldsymbol{v})\textcolor{blue}{-\big(2u^*-b^{-1}\big)f(\boldsymbol{u})h(\boldsymbol{u})},
\end{equation}
and
\begin{equation}\label{hf*-relation}
    \big((u+u^*)-(v-v^*)+b^{-1}\big)f^*(\boldsymbol{v})h(\boldsymbol{u})=
    \big((u-u^*)-(v-v^*)\big)\,h(\boldsymbol{u})f^*(\boldsymbol{v})
    \textcolor{blue}{+\big(2u^*+b^{-1}\big)f^*(\boldsymbol{u})h(\boldsymbol{u})}.
\end{equation}
The terms shown in \textcolor{blue}{blue} are called local terms, since they depend only on $\boldsymbol{u}$, but not on $\boldsymbol{v}$. 

Consider for example the relation \eqref{he-relation}, set $v-v^*=u-u^*$ and multiply it by $h^{-1}(\boldsymbol{u})$ from the left. Then the first term in the r.h.s. of  \eqref{he-relation} vanishes and one obtains 
\begin{equation}\label{e-holom}
    e(v+v^*,u-u^*)=e(u+u^*,u-u^*)\implies e=e(u-u^{*}).
\end{equation}
Similarly one finds 
\begin{equation}\label{f-holom}
    f=f(u+u^*),\quad
    e^*=e^*(u+u^*),\quad\text{and}\quad 
    f^*=f^*(u-u^*).
\end{equation}

The relations similar to \eqref{e-holom}--\eqref{f-holom} persist at higher levels as well. Let us consider the basic currents at level $1$
\begin{equation}
\begin{gathered}
    a(\boldsymbol{u})\overset{\text{def}}{=}h^{-1}(\boldsymbol{u})\langle\varnothing|\mathcal{L}(\boldsymbol{u})a_{-1}|\varnothing\rangle,\quad
    a^*(\boldsymbol{u})\overset{\text{def}}{=}\frac{1}{2u^*-b^{-1}}
    h^{-1}(\boldsymbol{u})\langle\varnothing|\mathcal{L}(\boldsymbol{u})a^*_{-1}|\varnothing\rangle,\\
    k(\boldsymbol{u})\overset{\text{def}}{=}\frac{1}{2u^*-b^{-1}}
    h^{-1}(\boldsymbol{u})\langle\varnothing|\mathcal{L}(\boldsymbol{u})\psi_{-\frac{1}{2}}\psi^*_{-\frac{1}{2}}|\varnothing\rangle,
\end{gathered}
\end{equation}
and their linear combinations 
\begin{equation}\label{rho-barrho-def}
    \rho(\boldsymbol{u})\overset{\text{def}}{=}a(\boldsymbol{u})-(2u^*-b^{-1})a^*(\boldsymbol{u})\quad\text{and}\quad
    \rho^*(\boldsymbol{u})\overset{\text{def}}{=}a(\boldsymbol{u})+(2u^*-2b-b^{-1})a^*(\boldsymbol{u})+2ik(\boldsymbol{u}).
\end{equation}
Using commutation relations \eqref{h-rho-relation} and \eqref{h-rhobar-relation} presented in the appendix, one finds
\begin{equation}\label{rho-relation}
    \rho=\rho\left(u-u^*\right),\quad
    \rho^*=\rho^*\left(u+u^*\right)
\end{equation}
Also the currents at level $-1$ 
\begin{equation}
\begin{gathered}
    \bar{a}(\boldsymbol{u})\overset{\text{def}}{=}\langle\varnothing|a_{1}\mathcal{L}(\boldsymbol{u})|\varnothing\rangle h^{-1}(\boldsymbol{u}),\quad
     \bar{a}^*(\boldsymbol{u})\overset{\text{def}}{=}
     \frac{1}{2u^*+b^{-1}}
     \langle\varnothing|a_{1}^*\mathcal{L}(\boldsymbol{u})|\varnothing\rangle h^{-1}(\boldsymbol{u}),\\
    \bar{k}(\boldsymbol{u})\overset{\text{def}}{=}
    \frac{1}{2u^*+b^{-1}}
    \langle\varnothing|\psi_{\frac{1}{2}}\psi^*_{\frac{1}{2}}\mathcal{L}(\boldsymbol{u})|\varnothing\rangle h^{-1}(\boldsymbol{u})
\end{gathered}
\end{equation}
and 
\begin{equation}\label{mu-barmu-def}
    \mu(\boldsymbol{u})\overset{\text{def}}{=}\bar{a}(\boldsymbol{u})+
    \left(2u^*+b^{-1}\right)\bar{a}^*(\boldsymbol{u})\quad\text{and}\quad
    \mu^*(\boldsymbol{u})\overset{\text{def}}{=}\bar{a}(\boldsymbol{u})-
    \left(2u^*-2b+b^{-1}\right)\bar{a}^*(\boldsymbol{u})-2i\bar{k}(\boldsymbol{u}),
\end{equation}
satisfy similar relations with $h(\boldsymbol{u})$, which imply
\begin{equation}\label{mu-relation}
    \mu=\mu\left(u+u^*\right),\quad
    \mu^*=\mu^*\left(u-u^*\right).
\end{equation}

In addition to \eqref{rho-relation} and \eqref{mu-relation}  there are composite currents, which also depend on one parameter instead of two. Namely, two level $1$ currents
\begin{equation}\label{sigma}
\begin{aligned}
    &\sigma(\boldsymbol{u})\overset{\text{def}}{=}\left(2u^*-b^{-1}\right)
    \left(e^*(\boldsymbol{u})e(\boldsymbol{u})+k(\boldsymbol{u})\right):\quad
    \sigma=\sigma(u-u^*),\\
    &\sigma^*(\boldsymbol{u})\overset{\text{def}}{=}
    (2u^*+b^{-1})\big(e(\boldsymbol{u})e^*(\boldsymbol{u})-k(\boldsymbol{u})\big)-2ia^*(\boldsymbol{u}):\quad
    \sigma^*=\sigma^*(u+u^*),
\end{aligned}
\end{equation}
and two level $-1$ currents
\begin{equation}\label{nu}
\begin{aligned}
    &\nu(\boldsymbol{u})\overset{\text{def}}{=}\left(2u^*+b^{-1}\right)
    \left(f(\boldsymbol{u})f^*(\boldsymbol{u})-\bar{k}(\boldsymbol{u})\right):\quad
    \nu=\nu(u+u^*),\\
    &\nu^*(\boldsymbol{u})\overset{\text{def}}{=}
    (2u^*-b^{-1})(f^*(\boldsymbol{u})f(\boldsymbol{u})+\bar{k}(\boldsymbol{u}))-2i\bar{a}^*(\boldsymbol{u}):\quad
    \nu^*=\nu^*(u-u^*).
\end{aligned}
\end{equation}
The derivation of \eqref{sigma} and \eqref{nu} is straightforward, yet tedious task and is based on commutation relation with $h(\boldsymbol{u})$. 

We will also need the following currents
\begin{equation}
    \uppsi(\boldsymbol{u})\overset{\text{def}}{=}
        \langle\varnothing\psi^*_{\frac{1}{2}}
    \mathcal{L}(\boldsymbol{u})\psi_{-\frac{1}{2}}|\varnothing\rangle h^{-1}(\boldsymbol{u})-\langle\varnothing|\psi^*_{\frac{1}{2}}
    \mathcal{L}(\boldsymbol{u})|\varnothing\rangle h^{-1}(\boldsymbol{u})
    \langle\varnothing|\mathcal{L}(\boldsymbol{u})\psi_{-\frac{1}{2}}|\varnothing\rangle h^{-1}(\boldsymbol{u}),
\end{equation}
and
\begin{equation}\label{psi-star-def}
    \uppsi^*(\boldsymbol{u})\overset{\text{def}}{=}\langle\varnothing|\psi_{\frac{1}{2}}
    \mathcal{L}(\boldsymbol{u})|\varnothing\rangle h^{-1}(\boldsymbol{u})
    \langle\varnothing|\mathcal{L}(\boldsymbol{u})\psi^*_{-\frac{1}{2}}|\varnothing\rangle h^{-1}(\boldsymbol{u})-
    \langle\varnothing|\psi_{\frac{1}{2}}
    \mathcal{L}(\boldsymbol{u})\psi^*_{-\frac{1}{2}}|\varnothing\rangle h^{-1}(\boldsymbol{u}),
\end{equation}
which can be shown to satisfy
\begin{equation}
    \uppsi(\boldsymbol{u})=\uppsi(u-u^*)\quad\text{and}\quad
    \uppsi^*(\boldsymbol{u})=\uppsi(u+u^*).
\end{equation}
\section{Current relations}\label{current-relations}
We define the parameters 
\begin{equation}
    h_1\overset{\text{def}}{=}\frac{b}{2}+b^{-1},\quad 
    h_2\overset{\text{def}}{=}\frac{b}{2}-b^{-1}
\end{equation}
and the currents
\begin{equation}
    e_1(u)\overset{\text{def}}{=}e(u),\quad
    e_2(u)\overset{\text{def}}{=}e^*\left(u+\frac{b}{2}\right),\quad
    f_1(u)\overset{\text{def}}{=}f^*(u)\quad\text{and}\quad
    f_2(u)\overset{\text{def}}{=}f\left(u+\frac{b}{2}\right),
\end{equation}
\begin{equation}
    \uppsi_1(u)\overset{\text{def}}{=}\uppsi(u),\quad
    \uppsi_2(u)\overset{\text{def}}{=}\uppsi^*(u)
\end{equation}

It is also convenient to define the currents labeled by pyramid partitions (see \eqref{rho-barrho-def}, \eqref{mu-barmu-def}, \eqref{sigma} and \eqref{nu})\footnote{The analog of MacMahon for $Y(\widehat{\mathfrak{gl}}(1|1))$ is naturally associated with the space of pyramid partitions (see \cite{Szendroi:2007nu}). It induces natural notations for the currents. There are two pyramid partitions with one stone which correspond to $e_1(u)$ and $e_2(u)$
\begin{equation}
  e_1(u)=e_{\includegraphics[scale=0.05]{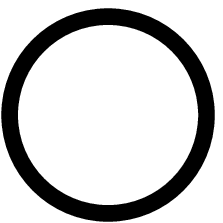}}(u)\quad\text{and}\quad
  e_2(u)=e_{\includegraphics[scale=0.05]{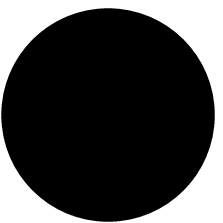}}(u).
\end{equation}
Also there are $4$ pyramid partitions with $2$ stones which correspond to the currents \eqref{pyramid-prtitions-two-stones}. Similar identification holds for $f$ currents as well.
}
\begin{equation}\label{pyramid-prtitions-two-stones}
\begin{gathered}
    e_{\includegraphics[scale=0.05]{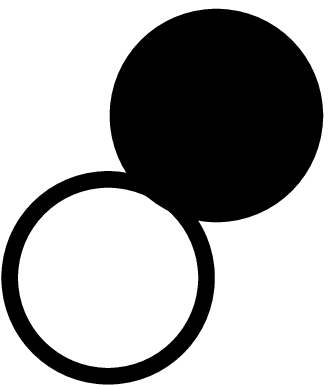}}(u)\overset{\text{def}}{=}\frac{i\rho(u)}{2h_2},\quad
    e_{\includegraphics[scale=0.05]{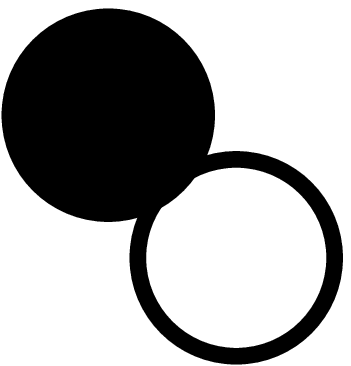}}(u)\overset{\text{def}}{=}\sigma(u)-\frac{i\rho(u)}{2h_2},\\
    e_{\includegraphics[scale=0.05]{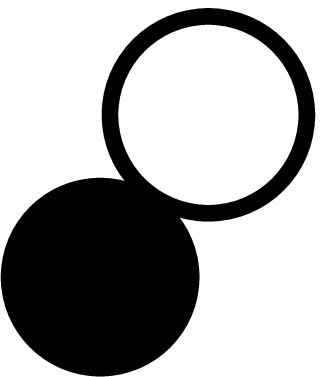}}(u)\overset{\text{def}}{=}
    \sigma^*\left(u+\frac{b}{2}\right)-\frac{i\rho^*\left(u+\frac{b}{2}\right)}{2h_1},\quad
    e_{\includegraphics[scale=0.05]{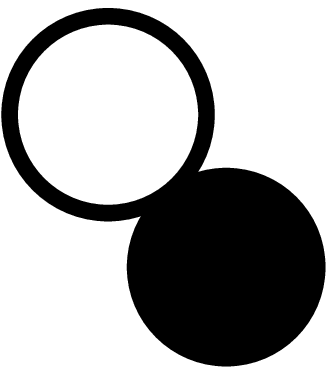}}(u)\overset{\text{def}}{=}\frac{i\rho^*\left(u+\frac{b}{2}\right)}{2h_1},
\end{gathered}
\end{equation}
and
\begin{equation}\label{pyramid-prtitions-two-stones-f}
\begin{gathered}
f_{\includegraphics[scale=0.05]{4b.eps}}(u)\overset{\text{def}}{=}\frac{i\mu\left(u+\frac{b}{2}\right)}{2h_2},\quad
f_{\includegraphics[scale=0.05]{3b.eps}}(u)\overset{\text{def}}{=}\nu\left(u+\frac{b}{2}\right)-\frac{i\mu\left(u+\frac{b}{2}\right)}{2h_2},\\
f_{\includegraphics[scale=0.05]{2b.eps}}(u)\overset{\text{def}}{=}\nu^*(u)-\frac{i\mu^*(u)}{2h_1},\quad
f_{\includegraphics[scale=0.05]{1b.eps}}(u)\overset{\text{def}}{=}\frac{i\mu^*(u)}{2h_1}.    
\end{gathered}
\end{equation}

One can derive the following  commutation relations (see appendix \ref{some-commutation-relations} for  details)
\begin{equation}\label{Yangian-relation-1}
   \left[\uppsi_i(u),\uppsi_j(v)\right]=0,\quad
   \left\{e_i(u),e_i(v)\right\}=0,\quad\left\{f_i(u),f_i(v)\right\}=0,
\end{equation}
\begin{equation}\label{ef+fe}
    \left\{e_i(u),f_j(v)\right\}=\delta_{ij}\frac{\uppsi_i(u)-\uppsi_i(v)}{u-v},
\end{equation}
\begin{multline}\label{ee-relation}
    (u-v-h_1)(u-v+h_1)\left(e_1(u)e_2(v)\textcolor{blue}{+\frac{e_{\includegraphics[scale=0.05]{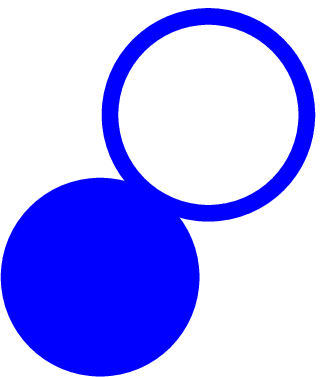}}(v)}{u-v-h_1}+\frac{e_{\includegraphics[scale=0.05]{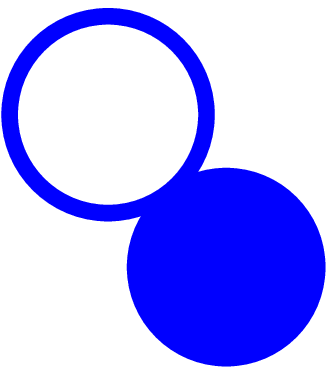}}(v)}{u-v+h_1}}\right)=\\=-
    (v-u-h_2)(v-u+h_2)\left(e_2(v)e_1(u)\textcolor{blue}{+\frac{e_{\includegraphics[scale=0.05]{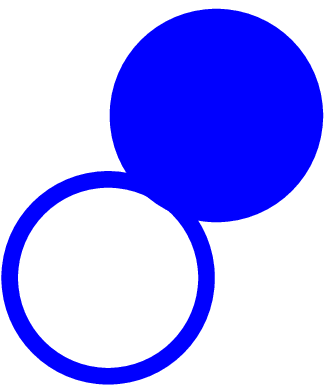}}(u)}{v-u-h_2}+\frac{e_{\includegraphics[scale=0.05]{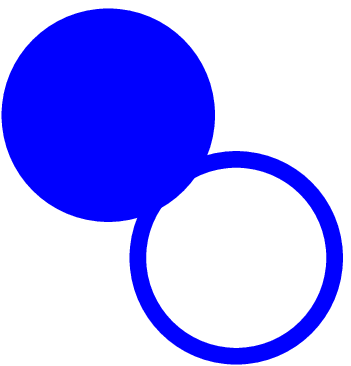}}(u)}{v-u+h_2}}\right)
\end{multline}
\begin{multline}\label{ff-relation}
    (u-v-h_1)(u-v+h_1)\left(f_2(u)f_1(v)\textcolor{blue}{+\frac{f_{\includegraphics[scale=0.05]{2.eps}}(u)}{v-u-h_1}+\frac{f_{\includegraphics[scale=0.05]{1.eps}}(u)}{v-u+h_1}}\right)=\\=-
    (v-u-h_2)(v-u+h_2)\left(f_1(v)f_2(u)\textcolor{blue}{+\frac{f_{\includegraphics[scale=0.05]{4.eps}}(u)}{u-v-h_2}+\frac{f_{\includegraphics[scale=0.05]{3.eps}}(v)}{u-v+h_2}}\right)
\end{multline}
\begin{multline}\label{psie-relation}
    \uppsi_i(u)e_j(v)=\frac{(u-v-h_j)(u-v+h_j)}{(u-v-h_i)(u-v+h_i)}e_j(v)\uppsi_i(u)+
    \\
    \textcolor{blue}{+\frac{h_i^2-h_j^2}{2h_i(u-v-h_i)}e_j(u-h_i)\uppsi_i(u)-\frac{h_i^2-h_j^2}{2h_i(u-v+h_i)}e_j(u+h_i)\uppsi_i(u)}
\end{multline}
\begin{multline}\label{psif-relation}
    \uppsi_i(u)f_j(v)=\frac{(u-v-h_i)(u-v+h_i)}{(u-v-h_j)(u-v+h_j)}f_j(v)\uppsi_i(u)+
    \\
    \textcolor{blue}{+\frac{h_j^2-h_i^2}{2h_j(u-v-h_j)}f_j(u-h_j)\uppsi_i(u)-\frac{h_j^2-h_i^2}{2h_j(u-v+h_j)}f_j(u+h_j)\uppsi_i(u)}
\end{multline}

The relations \eqref{Yangian-relation-1}--\eqref{psif-relation} should be understood in terms of modes of basic currents expanded at $u=\infty$
\begin{equation}
    e_k(u)=\frac{e_k^{(0)}}{u}+\frac{e_k^{(1)}}{u^2}+\dots,\quad
    f_k(u)=\frac{f_k^{(0)}}{u}+\frac{f_k^{(1)}}{u^2}+\dots,\quad
    \uppsi_k(u)=1+\frac{\uppsi_k^{(0)}}{u}+\frac{\uppsi_k^{(1)}}{u^2}+\dots
\end{equation}
We note that local terms in \eqref{Yangian-relation-1}--\eqref{psif-relation} depend only on one spectral parameter and do not contribute to commutation relations for modes with $(i,j)>1$. The relations with either $i\leq1$ or $j\leq1$ can be used to express higher currents in terms of basic ones. In particular, one finds from \eqref{ee-relation}
\begin{equation}
    \begin{aligned}
        &e_{\includegraphics[scale=0.05]{2b.eps}}(u)=-\frac{1}{2h_1}\left(\big\{e_1^{(0)},e_2^{(0)}\big\}-(u-h_1)\big\{e_1^{(0)},e_2(u)\big\}+\big\{e_1^{(1)},e_2(u)\big\}\right),\\
        &e_{\includegraphics[scale=0.05]{1b.eps}}(u)=\frac{1}{2h_1}\left(\big\{e_1^{(0)},e_2^{(0)}\big\}-(u+h_1)\big\{e_1^{(0)},e_2(u)\big\}+\big\{e_1^{(1)},e_2(u)\big\}\right),\\
        &e_{\includegraphics[scale=0.05]{4b.eps}}(u)=-\frac{1}{2h_2}\left(\big\{e_2^{(0)},e_1^{(0)}\big\}-(u-h_2)\big\{e_2^{(0)},e_1(u)\big\}+\big\{e_2^{(1)},e_1(u)\big\}\right),\\
        &e_{\includegraphics[scale=0.05]{3b.eps}}(u)=\frac{1}{2h_2}\left(\big\{e_2^{(0)},e_1^{(0)}\big\}-(u+h_2)\big\{e_2^{(0)},e_1(u)\big\}+\big\{e_2^{(1)},e_1(u)\big\}\right).
    \end{aligned}
\end{equation}
Similar relations hold for higher currents \eqref{pyramid-prtitions-two-stones-f} as well.

We note that \eqref{Yangian-relation-1}--\eqref{psif-relation} coincide exactly with commutation relations for $Y(\widehat{\mathfrak{gl}}(1|1))$ listed in \cite{Li:2020rij}. It is important to note that $RLL$-algebra contains more currents than $Y(\widehat{\mathfrak{gl}}(1|1))$. In particular, it has the current $h(\boldsymbol{u})$, which is not in $Y(\widehat{\mathfrak{gl}}(1|1))$. Moreover this current depends on two spectral parameters rather than one. The precise relation between two algebras, $RLL$ and $Y(\widehat{\mathfrak{gl}}(1|1))$, should be described by studying the center of the first along the lines of \cite{Litvinov:2020zeq}. We postpone this study to future publication.  

Also the algebra $Y(\widehat{\mathfrak{gl}}(1|1))$ contains Serre relations. Their derivation from $RLL$ algebra is straightforward task, which we expect to return in future publication (see \cite{Litvinov:2020zeq} for similar derivation for $Y(\widehat{\mathfrak{gl}}(1))$). 
\section{Conclusion}\label{concl}
In this paper we derived current relations of $Y(\widehat{\mathfrak{gl}}(1|1))$ \eqref{Yangian-relation-1}--\eqref{psif-relation} starting from $RLL$ realization. First, we defined the $R$-matrix, which serves as an intertwining operator for certain $\mathcal{N}=2$ superconformal algebra. The remarkable property of this $R-$matrix is that it depends on $4$ spectral parameters instead of $2$.  However, we have shown that the relations of $RLL$ algebra imply that some of the currents depend only on one spectral parameter. In particular, we defined the currents $e_i(u)$, $f_i(u)$ and $\uppsi_i(u)$ which satisfy the relations \eqref{Yangian-relation-1}--\eqref{psif-relation}.

There are several open questions, which we plan to address in future publications:
\begin{itemize}
    \item The relations \eqref{Yangian-relation-1}--\eqref{psif-relation} should be supplemented by Serre relations to define the correct algebra. The derivation of these relations starting from $RLL$ algebra is rather tedious. It is expected that Serre relations should be of quartic order of total charge $0$ (see e.g. \cite{Galakhov:2021xum}).  This requires studying of the $R-$matrix acting in $39$-dimensional space (as can be easily seen from the character formula \eqref{character}).
    \item It is interesting to describe explicitly the embedding $Y(\widehat{\mathfrak{gl}}(1))\oplus Y(\widehat{\mathfrak{gl}}(1))\in Y(\widehat{\mathfrak{gl}}(1|1))$, which is suggested by the results of \cite{Gaberdiel:2017hcn,Gaberdiel:2018nbs}. Namely, one expects to construct  two copies of  $Y(\widehat{\mathfrak{gl}}(1))$ currents $(e(u),f(u),\uppsi(u))$ and $(\hat{e}(u),\hat{f}(u),\hat{\uppsi}(u))$, as well as auxiliary fermionic currents $x(u)$, $\bar{x}(u)$, $y(u)$ and $\bar{y}(u)$  
    (see \cite{Gaberdiel:2017hcn,Gaberdiel:2018nbs}).
    \item Using current realization of $Y(\widehat{\mathfrak{gl}}(1|1))$ it is straightforward to derive Bethe ansatz equations following the lines of \cite{Litvinov:2020zeq,Chistyakova:2021yyd}. These Bethe ansatz equations will correspond in particular to quantum version of $\mathcal{N}=2$ KdV hierarchy and its $W_n$ generalizations. We plan to provide more details elsewhere.
\end{itemize}
\section*{Acknowledgements}
We acknowledge the discussions with Mikhail Bershtein, Boris Feigin, Mikhail Lashkevich, Andrii Liashyk and Sergei Parkhomenko.
\paragraph{Funding information:}
this work has been supported by the Russian Science Foundation under the grant 22-22-00991 and by Basis foundation. 
\Appendix
\section{Current commutation relations}\label{some-commutation-relations}
Here we provide some details on derivation of commutation relations of $Y(\widehat{\mathfrak{gl}}(1|1))$.
\paragraph{$h\rho$ and $h\rho^*$ commutation relations.} Using $RLL$ relations on level $1$ and charge $0$ one finds:
\begin{multline}\label{h-rho-relation}
    h(\boldsymbol{u})\rho(\boldsymbol{v})=\dfrac{((u-u^*)-(v-v^*))((u+u^*)-(v-v^*)-b+b^{-1})}{((u-u^*)-(v-v^*)-b+2b^{-1})((u+u^*)-(v-v^*)+b^{-1})}\rho(\boldsymbol{v})h(\boldsymbol{u})+\\+\dfrac{2(u^*+\frac{1}{2b})}{(u+u^*)-(v-v^*)+b^{-1}}h(\boldsymbol{u})\rho(\boldsymbol{u})-\\-
    \dfrac{2((u-u^*)-(v-v^*))(u^*-\frac{1}{2b})}{((u-u^*)-(v-v^*)-b+2b^{-1})((u+u^*)-(v-v^*)+b^{-1})}h(\boldsymbol{u})\rho^*(\boldsymbol{u})+
    \\+\frac{4((u-u^*)-(v-v^*))(u^*-\frac{1}{2b})}{((u-u^*)-(v-v^*)-b+2b^{-1})((u+u^*)-(v-v^*)+b^{-1})} h(\boldsymbol{u})a^*(\boldsymbol{u})+\\+\frac{4i((u-u^*)-(v-v^*))(u^*-\frac{1}{2b})}{((u-u^*)-(v-v^*)-b+2b^{-1})((u+u^*)-(v-v^*)+b^{-1})}e(\boldsymbol{v})h(\boldsymbol{u})e^*(\boldsymbol{u})
\end{multline}
and
\begin{multline}\label{h-rhobar-relation}
     h(\boldsymbol{u})\rho^*(\boldsymbol{v})=\dfrac{((u+u^*)-(v+v^*))((u-u^*)-(v+v^*)+b+b^{-1})}{((u-u^*)-(v+v^*)+b^{-1})((u+u^*)-(v+v^*)+b+2b^{-1})}\rho^*(\boldsymbol{v}) h(\boldsymbol{u})-\\-\dfrac{2(u^*-\frac{1}{2b})}{(u-u^*)-(v+v^*)+b^{-1}} h(\boldsymbol{u})\rho^*(\boldsymbol{u})+\\+\dfrac{2((u+u^*)-(v+v^*))(u^*+\frac{1}{2b}))}{((u-u^*)-(v+v^*)+b^{-1})((u+u^*)-(v+v^*)+b+2b^{-1})}h(\boldsymbol{u})(a(\boldsymbol{u})+a^*(\boldsymbol{u}))-\\-\dfrac{4i((u+u^*)-(v+v^*))(u^*+\frac{1}{2b}))}{((u-u^*)-(v+v^*)+b^{-1})((u+u^*)-(v+v^*)+b+2b^{-1})}e^*(\boldsymbol{v}) h(\boldsymbol{u})e(\boldsymbol{u})
\end{multline}
\paragraph{$ee$ commutation relation.} We introduce the following notations
\begin{equation}
\begin{gathered}
    \mathcal{L}_{\varnothing\psi}(\boldsymbol{u})\overset{\text{def}}{=}\langle\varnothing|\mathcal{L}(\boldsymbol{u})\psi_{-\frac{1}{2}}|\varnothing\rangle,\quad  \mathcal{L}_{\varnothing\psi^*}(\boldsymbol{u})\overset{\text{def}}{=}\langle\varnothing|\mathcal{L}(\boldsymbol{u})\psi^*_{-\frac{1}{2}}|\varnothing\rangle,
    \\
    \mathcal{L}_{\psi\varnothing}(\boldsymbol{u})\overset{\text{def}}{=}\langle\varnothing|\psi_{\frac{1}{2}}\mathcal{L}(\boldsymbol{u})|\varnothing\rangle,\quad \mathcal{L}_{\psi^*\varnothing}(\boldsymbol{u})\overset{\text{def}}{=}\langle\varnothing|\psi^*_{\frac{1}{2}}\mathcal{L}(\boldsymbol{u})|\varnothing\rangle, \\
    \mathcal{L}_{\psi\psi^*}(\boldsymbol{u})\overset{\text{def}}{=}\langle\varnothing|\psi_{\frac{1}{2}}\mathcal{L}(\boldsymbol{u})\psi^*_{-\frac{1}{2}}|\varnothing\rangle.
\end{gathered}    
\end{equation}
Then equation
\begin{equation}
\langle \varnothing |\mathcal{R}_{12}(\boldsymbol{u},\boldsymbol{v})\mathcal{L}_1(\boldsymbol{u})\mathcal{L}_2(\boldsymbol{v}) (\psi_1)_{-\frac{1}{2}}(\psi_2)_{-\frac{1}{2}}|\varnothing\rangle=\langle \varnothing |\mathcal{L}_2(\boldsymbol{v})\mathcal{L}_1(\boldsymbol{u})\mathcal{R}_{12}(\boldsymbol{u},\boldsymbol{v})(\psi_1)_{-\frac{1}{2}}(\psi_2)_{-\frac{1}{2}}|\varnothing\rangle
\end{equation}
can be rewritten in the explicit form
\begin{equation}\label{ee-1}
    \mathcal{L}_{\varnothing\psi}(\boldsymbol{u})\mathcal{L}_{\varnothing\psi}(\boldsymbol{v})=-\dfrac{(u-u^*)-(v+v^*)-b^{-1}}{(u+u^*)-(v-v^*)+b^{-1}}\mathcal{L}_{\varnothing\psi}(\boldsymbol{v})\mathcal{L}_{\varnothing\psi}(\boldsymbol{u}),
\end{equation}
where we considered the action of $R$-matrix on $\mathcal{R}_{12}(\boldsymbol{u},\boldsymbol{v})(\psi_1)_{-\frac{1}{2}}(\psi_2)_{-\frac{1}{2}}|\varnothing\rangle$. It is important to note the presence of the "minus" sign in this equation due to taking into account the fermions.

Consider also the equation
\begin{equation}
\langle \varnothing |\mathcal{R}_{12}(\boldsymbol{u},\boldsymbol{v})\mathcal{L}_1(\boldsymbol{u})\mathcal{L}_2(\boldsymbol{v}) (\psi_1)_{-\frac{1}{2}}|\varnothing\rangle=\langle \varnothing |\mathcal{L}_2(\boldsymbol{v})\mathcal{L}_1(\boldsymbol{u})\mathcal{R}_{12}(\boldsymbol{u},\boldsymbol{v})(\psi_1)_{-\frac{1}{2}}|\varnothing\rangle,
\end{equation}
which takes the form (compare to \eqref{RLL-3})
\begin{equation}\label{ee-2}
     \mathcal{L}_{\varnothing\psi}(\boldsymbol{u})h(\boldsymbol{v})=\dfrac{(u+u^*)-(v+v^*)}{(u+u^*)-(v-v^*)+b^{-1}}h(\boldsymbol{v})\mathcal{L}_{\varnothing\psi}(\boldsymbol{u})+\dfrac{2v^*+b^{-1}}{(u+u^*)-(v-v^*)+b^{-1}}\mathcal{L}_{\varnothing\psi}(\boldsymbol{v})h(\boldsymbol{u}).
\end{equation}
Then multiplying (\ref{ee-1}) by $h^{-1}(\boldsymbol{u})h^{-1}(\boldsymbol{v})$ on the left and using (\ref{ee-2}), we get the anticommutation relation
\begin{equation}\label{anticommutator e(u)e(v)}
    \{e(\boldsymbol{u}),e(\boldsymbol{v})\}=\dfrac{2u^*+b^{-1}}{(u+u^*)-(v-v^*)+b^{-1}}e^2(\boldsymbol{u})-\dfrac{2v^*+b^{-1}}{(u-u^*)-(v+v^*)-b^{-1}}e^2(\boldsymbol{v}).
\end{equation}
The relations \eqref{anticommutator e(u)e(v)} should not have any poles. It implies that $e^2(\boldsymbol{u})=0$ and $\{e(\boldsymbol{u}),e(\boldsymbol{v})\}=0$.
\paragraph{$ee^*$ commutation relation.}
We consider the equation
\begin{equation}
\langle \varnothing |\mathcal{R}_{12}(\boldsymbol{u},\boldsymbol{v})\mathcal{L}_1(\boldsymbol{u})\mathcal{L}_2(\boldsymbol{v}) (\psi_1)_{-\frac{1}{2}}(\psi^*_2)_{-\frac{1}{2}}|\varnothing\rangle=\langle \varnothing |\mathcal{L}_2(\boldsymbol{v})\mathcal{L}_1(\boldsymbol{u})\mathcal{R}_{12}(\boldsymbol{u},\boldsymbol{v})(\psi_1)_{-\frac{1}{2}}(\psi^*_2)_{-\frac{1}{2}}|\varnothing\rangle.
\end{equation}
In principle, one can add to this formula an arbitrary linear combination of $RLL$-relations at level 1 with charge 0: 
\begin{multline}
    \langle \varnothing |\mathcal{R}_{12}(\boldsymbol{u},\boldsymbol{v})\mathcal{L}_1(\boldsymbol{u})\mathcal{L}_2(\boldsymbol{v}) (\psi_1)_{-\frac{1}{2}}(\psi^*_2)_{-\frac{1}{2}}|\varnothing\rangle+\alpha_1 \langle \varnothing |\mathcal{R}_{12}(\boldsymbol{u},\boldsymbol{v})\mathcal{L}_1(\boldsymbol{u})\mathcal{L}_2(\boldsymbol{v}) (a_1)_{-1}|\varnothing\rangle+\\+\alpha_2 \langle \varnothing |\mathcal{R}_{12}(\boldsymbol{u},\boldsymbol{v})\mathcal{L}_1(\boldsymbol{u})\mathcal{L}_2(\boldsymbol{v}) (\psi_1)_{-\frac{1}{2}}(\psi^*_1)_{-\frac{1}{2}}|\varnothing\rangle+\dots
\end{multline}
but we leave only those terms which, after multiplying by $h^{-1}(\boldsymbol{u})h^{-1}(\boldsymbol{v})$ from the left, turn out to be local:
\begin{multline}\label{lin comb RLL}
     \langle \varnothing |\mathcal{R}_{12}(\boldsymbol{u},\boldsymbol{v})\mathcal{L}_1(\boldsymbol{u})\mathcal{L}_2(\boldsymbol{v}) (\psi_1)_{-\frac{1}{2}}(\psi^*_2)_{-\frac{1}{2}}|\varnothing\rangle+\alpha \langle \varnothing |\mathcal{R}_{12}(\boldsymbol{u},\boldsymbol{v})\mathcal{L}_1(\boldsymbol{u})\mathcal{L}_2(\boldsymbol{v}) (a_2)_{-1}|\varnothing\rangle+\\+\beta\langle \varnothing |\mathcal{R}_{12}(\boldsymbol{u},\boldsymbol{v})\mathcal{L}_1(\boldsymbol{u})\mathcal{L}_2(\boldsymbol{v}) (a^*_2)_{-1}|\varnothing\rangle+\gamma\langle \varnothing |\mathcal{R}_{12}(\boldsymbol{u},\boldsymbol{v})\mathcal{L}_1(\boldsymbol{u})\mathcal{L}_2(\boldsymbol{v}) (\psi_2)_{-\frac{1}{2}}(\psi^*_2)_{-\frac{1}{2}}|\varnothing\rangle.
\end{multline}
The coefficients $\alpha$, $\beta$, $\gamma$ are found from the condition of cancellation of the terms
\begin{equation}
       \langle \varnothing |\mathcal{L}_2(\boldsymbol{v})\mathcal{L}_1(\boldsymbol{u})(a_2)_{-1}|\varnothing\rangle, \quad \langle \varnothing |\mathcal{L}_2(\boldsymbol{v})\mathcal{L}_1(\boldsymbol{u})(a^*_2)_{-1}|\varnothing\rangle,\quad
       \langle \varnothing |\mathcal{L}_2(\boldsymbol{v})\mathcal{L}_1(\boldsymbol{u})(\psi_2)_{-\frac{1}{2}}(\psi^*_2)_{-\frac{1}{2}}|\varnothing\rangle,
\end{equation}
in the right side of the $RLL$-relation (\ref{lin comb RLL}). Their explicit expressions are
\begin{equation}
\begin{gathered}
    \alpha=-\dfrac{i(2v^*-b^{-1})}{((u-u^*)-(v-v^*))(\Delta+h_1+\frac{b}{2}))},\quad
    \beta=-\dfrac{2i((u-u^*)-v+\frac{1}{2b})}{((u-u^*)-(v-v^*))(\Delta+h_1+\frac{b}{2}))},\\
    \gamma=\dfrac{2}{((u-u^*)-(v-v^*))(\Delta+h_1+\frac{b}{2}))}-\dfrac{2v^*+b^{-1}}{(u-u^*)-(v-v^*)}.
\end{gathered}
\end{equation}
Here and below, we use the following notation 
\begin{equation}
    \Delta=(u-u^*)-(v+v^*),\quad  
    h_1=\dfrac{b}{2}+b^{-1},\quad
    h_2=\dfrac{b}{2}-b^{-1}.
\end{equation}

Then, taking into account the action of the $R$-matrix on the states of level $1$ with charge $0$, we get
\begin{multline}
    \mathcal{L}_{\varnothing\psi}(\boldsymbol{u})\mathcal{L}_{\varnothing\psi^*}(\boldsymbol{v})+\alpha h(\boldsymbol{u}) \mathcal{L}_{\varnothing a}(\boldsymbol{v})+\beta h(\boldsymbol{u}) \mathcal{L}_{\varnothing a^*}(\boldsymbol{v})+\gamma \mathcal{L}_{\varnothing k}(\boldsymbol{v})=\\=-\dfrac{((u+u^*)-(v+v^*))(\Delta+h_2+\frac{b}{2})}{((u-u^*)-(v-v^*))(\Delta+h_1+\frac{b}{2})} \mathcal{L}_{\varnothing\psi^*}(\boldsymbol{v})\mathcal{L}_{\varnothing\psi}(\boldsymbol{u})-\\-\dfrac{(2v^*-b^{-1})(\Delta+h_2+\frac{b}{2})}{((u-u^*)-(v-v^*))(\Delta+h_1+\frac{b}{2})}h(\boldsymbol{v})\mathcal{L}_{\varnothing k}(\boldsymbol{u})+\\+\dfrac{i(2v^*-b^{-1})}{((u-u^*)-(v-v^*))(\Delta+h_1+\frac{b}{2})}h(\boldsymbol{v})(\mathcal{L}_{\varnothing a^*}(\boldsymbol{u})-\mathcal{L}_{\varnothing a}(\boldsymbol{u})).
\end{multline}
Multiplying this expression by $h^{-1}(\boldsymbol{u})h^{-1}(\boldsymbol{v})$ on the left and using commutation relations $h^{-1}(\boldsymbol{v})\mathcal{L}_{\varnothing\psi}(\boldsymbol{u})$ and $h^{-1}(\boldsymbol{u})\mathcal{L}_{\varnothing\psi^*}(\boldsymbol{v})$, one obtains
\begin{multline}\label{ee*-relation}
    e(\boldsymbol{u})e^*(\boldsymbol{v})\textcolor{blue}{-\frac{i\rho^*(\boldsymbol{v})}{(\Delta+h_1+\frac{b}{2})(\Delta-h_1+\frac{b}{2})}}\textcolor{blue}{+\frac{\overbrace{(2v^*+b^{-1})\big(e(\boldsymbol{v})e^*(\boldsymbol{v})-k(\boldsymbol{v})\big)-2ia^*(\boldsymbol{v})}^{\sigma^*(\boldsymbol{v})}}
    {(\Delta-h_1+\frac{b}{2})}}=\\=
    -\dfrac{(\Delta+h_2+\frac{b}{2})(\Delta-h_2+\frac{b}{2})}{(\Delta+h_1+\frac{b}{2})(\Delta-h_1+\frac{b}{2})}e^*(\boldsymbol{v})e(\boldsymbol{u})\textcolor{blue}{-\frac{i\rho(\boldsymbol{u})}{(\Delta+h_1+\frac{b}{2})(\Delta-h_1+\frac{b}{2})}}\textcolor{blue}{-}\\
    \textcolor{blue}
    {-\frac{(\Delta+h_2+\frac{b}{2})}{(\Delta+h_1+\frac{b}{2})(\Delta-h_1+\frac{b}{2})}
    \underbrace{(2u^*-b^{-1})\left(e^*(\boldsymbol{u})e(\boldsymbol{u})+k(\boldsymbol{u})\right)}_{\sigma(\boldsymbol{u})}}
    ,
\end{multline}
We see that, this commutation relation corresponds to \eqref{ee-relation}. 
\paragraph{$e^*f$ commutation relation.}
Let us consider the equation
\begin{equation}
\langle \varnothing |(\psi_2)_{\frac{1}{2}}\mathcal{R}_{12}(\boldsymbol{u},\boldsymbol{v})\mathcal{L}_1(\boldsymbol{u})\mathcal{L}_2(\boldsymbol{v}) (\psi_1^*)_{-\frac{1}{2}}|\varnothing\rangle=\langle \varnothing |(\psi_2)_{\frac{1}{2}}\mathcal{L}_2(\boldsymbol{v})\mathcal{L}_1(\boldsymbol{u})\mathcal{R}_{12}(\boldsymbol{u},\boldsymbol{v})(\psi_1^*)_{-\frac{1}{2}}|\varnothing\rangle.
\end{equation}
Then from the definition of the $R$-matrix one finds
\begin{multline}\label{e^*f-first eq}
    -\dfrac{2u^*-b^{-1}}{(u-u^*)-(v+v^*)+b^{-1}}\mathcal{L}_{\psi\psi^*}(\boldsymbol{u})h(\boldsymbol{v})-\dfrac{(u+u^*)-(v+v^*)}{(u-u^*)-(v+v^*)+b^{-1}}\mathcal{L}_{\varnothing\psi^*}(\boldsymbol{u})\mathcal{L}_{\psi\varnothing}(\boldsymbol{v})=\\=\dfrac{(u-u^*)-(v-v^*)}{(u-u^*)-(v+v^*)+b^{-1}}\mathcal{L}_{\psi\varnothing}(\boldsymbol{v})\mathcal{L}_{\varnothing\psi^*}(\boldsymbol{u})-\dfrac{2u^*-b^{-1}}{(u-u^*)-(v+v^*)+b^{-1}}\mathcal{L}_{\psi\psi^*}(\boldsymbol{v})h(\boldsymbol{u}).
\end{multline}
Let us also consider two auxiliary equations
\begin{align}
\langle \varnothing |\mathcal{R}_{12}(\boldsymbol{u},\boldsymbol{v})\mathcal{L}_1(\boldsymbol{u})\mathcal{L}_2(\boldsymbol{v}) (\psi_1^*)_{-\frac{1}{2}}|\varnothing\rangle=\langle \varnothing |\mathcal{L}_2(\boldsymbol{v})\mathcal{L}_1(\boldsymbol{u})\mathcal{R}_{12}(\boldsymbol{u},\boldsymbol{v})(\psi_1^*)_{-\frac{1}{2}}|\varnothing\rangle, \\
\langle \varnothing |(\psi_2)_{\frac{1}{2}}\mathcal{R}_{12}(\boldsymbol{u},\boldsymbol{v})\mathcal{L}_1(\boldsymbol{u})\mathcal{L}_2(\boldsymbol{v}) |\varnothing\rangle=\langle \varnothing |(\psi_2)_{\frac{1}{2}}\mathcal{L}_2(\boldsymbol{v})\mathcal{L}_1(\boldsymbol{u})\mathcal{R}_{12}(\boldsymbol{u},\boldsymbol{v})|\varnothing\rangle,
\end{align}
which can be rewritten explicitly as 
\begin{equation}
    \mathcal{L}_{\varnothing\psi^*}(\boldsymbol{u})h(\boldsymbol{v})=\dfrac{(u-u^*)-(v-v^*)}{(u-u^*)-(v+v^*)+b^{-1}}h(\boldsymbol{v})\mathcal{L}_{\varnothing\psi^*}(\boldsymbol{u})-\dfrac{2u^*-b^{-1}}{(u-u^*)-(v+v^*)+b^{-1}}\mathcal{L}_{\varnothing\psi^*}(\boldsymbol{v})h(\boldsymbol{u}),
\end{equation}
and
\begin{equation}
     \mathcal{L}_{\psi\varnothing}(\boldsymbol{v})h(\boldsymbol{u})=\dfrac{(u+u^*)-(v+v^*)}{(u-u^*)-(v+v^*)+b^{-1}}h(\boldsymbol{u})\mathcal{L}_{\psi\varnothing}(\boldsymbol{v})-\dfrac{2u^*-b^{-1}}{(u-u^*)-(v+v^*)+b^{-1}}\mathcal{L}_{\psi\varnothing}(\boldsymbol{u})h(\boldsymbol{v}),
\end{equation}
Multiplying (\ref{e^*f-first eq}) by $h^{-1}(\boldsymbol{u})h^{-1}(\boldsymbol{v})$ from the right and using these relations, one finds
\begin{multline}\label{e^*f-second eq}
     -\mathcal{L}_{\varnothing\psi^*}(\boldsymbol{u})h^{-1}(\boldsymbol{u})\mathcal{L}_{\psi\varnothing}(\boldsymbol{v})h^{-1}(\boldsymbol{v})-\dfrac{2u^*-b^{-1}}{(u-u^*)-(v+v^*)+b^{-1}}\mathcal{L}_{\psi\psi^*}(\boldsymbol{u})h^{-1}(\boldsymbol{u})-\\-\dfrac{2u^*-b^{-1}}{(u-u^*)-(v+v^*)+b^{-1}}\mathcal{L}_{\varnothing\psi^*}(\boldsymbol{u})h^{-1}(\boldsymbol{u})\mathcal{L}_{\psi\varnothing}(\boldsymbol{u})h^{-1}(\boldsymbol{u})=\\=\mathcal{L}_{\psi\varnothing}(\boldsymbol{v})h^{-1}(\boldsymbol{v})\mathcal{L}_{\varnothing\psi^*}(\boldsymbol{u})h^{-1}(\boldsymbol{u})-\dfrac{2u^*-b^{-1}}{(u-u^*)-(v+v^*)+b^{-1}}\mathcal{L}_{\psi\psi^*}(\boldsymbol{v})h^{-1}(\boldsymbol{v})+\\+\dfrac{2u^*-b^{-1}}{(u-u^*)-(v+v^*)+b^{-1}}\mathcal{L}_{\psi\varnothing}(\boldsymbol{v})h^{-1}(\boldsymbol{v})\mathcal{L}_{\varnothing\psi^*}(\boldsymbol{v})h^{-1}(\boldsymbol{v}).
\end{multline}
Equation (\ref{he*-relation}) implies
\begin{equation}
    e^*(u-u^*+b^{-1})=\dfrac{1}{2u^*-b^{-1}}\mathcal{L}_{\varnothing\psi^*}(\boldsymbol{u})h^{-1}(\boldsymbol{u}),
\end{equation}
from which one can obtain
\begin{equation}
    \{e^*(u-u^*+b^{-1}),f(\boldsymbol{v})\}=\dfrac{\uppsi^*(u,-u^*+b^{-1})-\uppsi^*(\boldsymbol{v})}{(u-u^*)-(v+v^*)+b^{-1}},
\end{equation}
where (compare to \eqref{psi-star-def})
\begin{equation}
    \uppsi^*(u+u^*)\overset{\text{def}}{=}\mathcal{L}_{\psi\varnothing}(\boldsymbol{u})h^{-1}(\boldsymbol{u})\mathcal{L}_{\varnothing\psi^*}(\boldsymbol{u})h^{-1}(\boldsymbol{u})-\mathcal{L}_{\psi\psi^*}(\boldsymbol{u})h^{-1}(\boldsymbol{u}).
\end{equation}
So, this formula corresponds to (\ref{ef+fe}). Similarly, one can get the commutation relation for $e(\boldsymbol{u})$ and $f^*(\boldsymbol{v})$:
\begin{equation}
    \{e(u+u^*+b^{-1}),f^*(\boldsymbol{v})\}=\dfrac{\uppsi(u,-u^*-b^{-1})-\uppsi(\boldsymbol{v})}{(u-u^*)-(v-v^*)}
\end{equation}
with
\begin{equation}
    \uppsi(u-u^*)\overset{\text{def}}{=}\mathcal{L}_{\psi^*\psi}(\boldsymbol{u})h^{-1}(\boldsymbol{u})-\mathcal{L}_{\psi^*\varnothing}(\boldsymbol{u})h^{-1}(\boldsymbol{u})\mathcal{L}_{\varnothing\psi}(\boldsymbol{u})h^{-1}(\boldsymbol{u}).
\end{equation}

The remaining relations from \eqref{Yangian-relation-1}--\eqref{psif-relation} can be derived in a similar way.

\bibliographystyle{MyStyle} 
\bibliography{MyBib}
\end{document}